\documentclass[sigconf, nonacm]{acmart}

\makeatletter
\@ifundefined{Bbbk}{}{}
\makeatother

\usepackage{amsmath,amssymb,mathtools}
\usepackage{amsthm}

\usepackage{booktabs,tabularx}

\usepackage{graphicx}
\usepackage{tikz}
\usepackage{booktabs}       
\usepackage{amsfonts}       
\usepackage{nicefrac}       
\usepackage{microtype}      
\usepackage{graphicx}
\usepackage{algorithm}
\usepackage{algorithmic} 
\usepackage{enumitem}
\usepackage{float}
\usepackage{multirow}
\usepackage[table]{xcolor} 
\usetikzlibrary{positioning,calc,arrows.meta} 
\definecolor{myblue}{RGB}{0,114,178}
\definecolor{mygreen}{RGB}{0,158,115}
\definecolor{myorange}{RGB}{213,94,0}

\AtBeginDocument{%
  }

\begin{document}

\title{TGSBM: Transformer-Guided Stochastic Block Model for Link Prediction}

\author{Zhejian Yang}
\orcid{0000-0002-6980-8062}
\affiliation{%
  \institution{Jilin University}
  \city{Changchun}
  \state{Jilin}
  \country{China}
}
\email{zjyang22@mails.jlu.edu.cn}

\author{Songwei Zhao}
\orcid{0000-0001-6149-3214}
\authornote{Corresponding authors: Songwei Zhao and Hechang Chen}
\affiliation{%
  \institution{Jilin University}
  \city{Changchun}
  \state{Jilin}
  \country{China}
}
\email{zhaosw22@mails.jlu.edu.cn}

\author{Zilin Zhao}
\orcid{0009-0003-2190-958X}
\affiliation{%
  \institution{Jilin University}
  \city{Changchun}
  \state{Jilin}
  \country{China}
}
\email{zhao_zilin@outlook.com}

\author{Hechang Chen}
\authornotemark[1]
\orcid{0000-0001-7835-9556}
\affiliation{%
  \institution{Jilin University}
  \city{Changchun}
  \state{Jilin}
  \country{China}
}
\email{chenhc@jlu.edu.cn}





\renewcommand{\shortauthors}{Zhejian Yang, Songwei Zhao, Zilin Zhao, and Hechang Chen}

\begin{abstract}
Link prediction is a cornerstone of the Web ecosystem, powering applications from recommendation and search to knowledge graph completion and collaboration forecasting. However, large-scale networks present unique challenges: they contain hundreds of thousands of nodes and edges with heterogeneous and overlapping community structures that evolve over time. Existing approaches face notable limitations: traditional graph neural networks struggle to capture global structural dependencies, while recent graph transformers achieve strong performance but incur quadratic complexity and lack interpretable latent structure.
We propose \textbf{TGSBM} (Transformer-Guided Stochastic Block Model), a framework that integrates the principled generative structure of Overlapping Stochastic Block Models with the representational power of sparse Graph Transformers. TGSBM comprises three main components: (i) \emph{expander-augmented sparse attention} that enables near-linear complexity and efficient global mixing, (ii) a \emph{neural variational encoder} that infers structured posteriors over community memberships and strengths, and (iii) a \emph{neural edge decoder} that reconstructs links via OSBM's generative process, preserving interpretability.
Experiments across diverse benchmarks demonstrate competitive performance (mean rank 1.6 under HeaRT protocol), superior scalability (up to $6\times$ faster training), and interpretable community structures. These results position TGSBM as a practical approach that strikes a balance between accuracy, efficiency, and transparency for large-scale link prediction.
\end{abstract}

\ccsdesc[500]{Computing methodologies~Latent variable models}
\ccsdesc[500]{Information systems~Data mining}

\keywords{Stochastic block model; Graph Transformer; Link prediction}


\maketitle

\section{Introduction}

Link prediction is fundamental to the Web ecosystem, supporting a diverse range of applications across multiple domains. From recommending professional contacts on LinkedIn to expanding knowledge graphs such as Wikidata, from suggesting citations in scholarly search engines to predicting collaborations on platforms like GitHub, accurate link forecasting directly impacts search, recommendation, and knowledge discovery in large-scale networks~\cite{newman2001clustering, singh2024social}. Beyond these platforms, link prediction is also central to Web security (e.g., detecting suspicious connections), social analysis, and biomedical discovery~\cite{hu2020open}, highlighting its broad societal and economic significance. However, real-world networks often contain hundreds of thousands of nodes and edges, and exhibit heterogeneous and overlapping community structures that evolve dynamically over time, making link prediction a challenging task. Consequently, classical heuristics and existing neural approaches are often insufficient, exposing the need for principled and scalable methods designed specifically for large-scale link prediction.

Current graph learning approaches face notable limitations when applied to large-scale networks. Traditional MPNNs such as GCN~\cite{kipf2016semi} and GraphSAGE~\cite{hamilton2017inductive} struggle to capture the global structural dependencies essential for understanding community-driven graphs. Pairwise-enhanced methods such as SEAL~\cite{zhang2018link} and NBFNet~\cite{zhu2021neural} improve expressiveness but require separate inference for each candidate edge, which is computationally prohibitive in networks with large numbers of potential links. Graph Transformers~\cite{ying2021transformers,wu2022nodeformer} offer strong representational power but often incur quadratic complexity, limiting scalability. Moreover, they typically behave as black boxes, obscuring the latent community structures behind link formation--an obstacle for platforms that require transparency in recommendation explanation and moderation. Therefore, developing models that can efficiently capture global graph structure while maintaining interpretable community representations is critical for advancing link prediction in real-world applications.

Despite recent progress, three important challenges remain to be addressed in large-scale link prediction. 
\textbf{Challenge 1: Efficiency.} Can we propagate information across large-scale graphs in near-linear complexity while maintaining low effective diameter for global mixing? 
\textbf{Challenge 2: Overlapping communities.} Real-world entities naturally belong to multiple groups--researchers active in several fields, users participating in diverse interest circles, or pages spanning multiple topics. How can we capture such overlapping memberships in a probabilistic yet interpretable manner? 
\textbf{Challenge 3: Interpretability.} Recommendation systems increasingly demand explainable predictions (e.g., explaining why a link was predicted based on shared community memberships). How can models retain interpretable latent structure without sacrificing predictive performance?

\begin{figure}[t]
  \centering
  \includegraphics[width=0.9\linewidth]{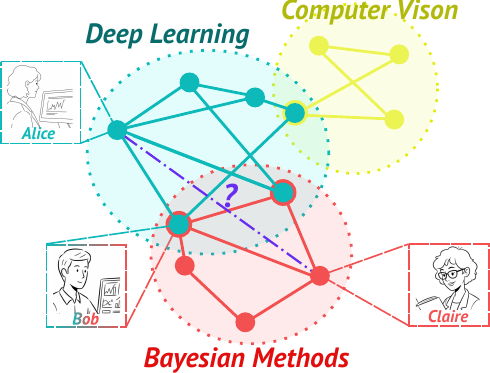}
  \caption{\textbf{Link prediction with overlapping communities.} Researchers belong to multiple overlapping communities (Deep Learning, Computer Vision, Bayesian Methods). The task is to predict missing links (dashed edge with ``?'') based on shared community memberships while preserving interpretability.}
  \Description{Collaboration network showing three researchers (Alice, Bob, Claire) connected by edges, with color-coded regions representing three overlapping research communities: Deep Learning (cyan), Computer Vision (yellow), and Bayesian Methods (red). A purple dashed edge with a question mark indicates the link prediction task.}
  \label{fig:intro}
\end{figure}

Figure~\ref{fig:intro} illustrates these challenges: in a collaboration network, researchers like Alice, Bob, and Claire belong to multiple overlapping communities (e.g., Deep Learning, Computer Vision, Bayesian Methods). Link prediction must determine whether a connection should exist between nodes (e.g., the purple dashed edge) based on their shared community memberships, requiring models that can efficiently capture global structure, represent overlapping affiliations, and provide interpretable explanations for predictions.

To address these challenges, we propose \textbf{TGSBM} (Transformer-Guided Stochastic Block Model), which integrates the interpretable latent structure of Overlapping Stochastic Block Models (OSBM)~\cite{latouche2011overlapping,miller2009nonparametric,zhu2016max} with the representational power of sparse Graph Transformers. We observe that these two paradigms are complementary: OSBM provides a principled generative model of community-driven link formation, while sparse Transformers enable efficient learning of complex patterns at scale. TGSBM addresses these challenges through three main components: (1) \emph{Expander-augmented sparse attention} that achieves efficient global mixing in $O((|E|+dN)H)$ complexity while preserving community structure; (2) a \emph{neural variational encoder} that outputs structured posteriors over OSBM latent variables, including binary memberships, continuous strengths, and stick-breaking priors; and (3) a \emph{neural edge decoder} that reconstructs links via the OSBM generative process, maintaining interpretability through probabilistic community interactions.

Our main contributions are threefold:
\begin{itemize}
\item We introduce Transformer-Guided Stochastic Block Model (TGSBM), a novel framework that unifies OSBM latent community modeling with sparse Graph Transformers, enabling interpretable and scalable link prediction on large-scale networks.  
\item We develop a principled design that combines expander-augmented sparse attention with a neural variational inference scheme, ensuring efficient global propagation while capturing overlapping community memberships in a probabilistic manner.  
\item We validate TGSBM across diverse benchmarks and evaluation protocols, including HeaRT~\cite{li2023evaluating}, demonstrating competitive performance, significant runtime advantages, and interpretable community structures that are beneficial for practical deployment.  
\end{itemize}


\begin{table*}[t]
\centering
\caption{Notation for TGSBM.}
\vspace{-1mm}
\label{tab:notation_compact}
\begin{tabular}{@{}l l p{4.9cm} | l l p{4.9cm}@{}}
\toprule
\multicolumn{3}{c}{\textbf{Graph \& Encoder}} & \multicolumn{3}{c}{\textbf{OSBM Latents \& Variational}} \\
\midrule
\textbf{Symbol} & \textbf{Shape} & \textbf{Meaning} & \textbf{Symbol} & \textbf{Shape} & \textbf{Meaning} \\
\midrule
$G=(\mathcal{V},\mathcal{E})$ & -- & Input undirected graph; $N=|\mathcal{V}|$ & $K$ & scalar & Truncation level (number of latent communities) \\
$\mathbf{A}$ & $\{0,1\}^{N\times N}$ & Observed adjacency matrix & $\mathbf{B}$ & $\{0,1\}^{N \times K}$ & Node membership matrix \\
$\mathbf{X}$ & $\mathbb{R}^{N\times F}$ & (Optional) node features & $\mathbf{R}$ & $\mathbb{R}^{N \times K}$ & Community strength matrix \\
$\mathcal{E}_{\text{attn}}$ & set & Attention edges: $\mathcal{E}_{\text{local}}\cup\mathcal{E}_{\exp}$ & $\mathbf{Z}$ & $\mathbb{R}^{N \times K}$ & Node latent embedding matrix \\
$d_{\exp}$ & scalar & Degree of expander overlay & $\mathbf{v}$ & $\mathbb{R}^{K}$ & Stick-breaking variables \\
$H, d_h, D$ & scalars & \#heads, head dim, model width & $\boldsymbol{\alpha}$ & $\mathbb{R}^{K}$ & Community prevalence parameters \\
$L$ & scalar & \#encoder layers & $\boldsymbol{\mu}_n,\boldsymbol{\sigma}_n$ & $\mathbb{R}^{K}$ & Mean/STD of $\mathbf{r}_n$ posterior \\
$f$ & map & Two-layer MLP: $\mathbb{R}^{K}\!\to\!\mathbb{R}^{d_{\mathrm{mlp}}}$ & $\tilde{W}$ & $\mathbb{R}^{(K+1)\times(K+1)}$ & Augmented OSBM interaction matrix \\
$W_\pi,W_\mu,W_\sigma$ & $\mathbb{R}^{K\times D}$ & Projections for variational heads & $\mathcal{S}$ & set & Sampled training edge-pairs \\
 &  &  & $\tau$ & scalar & Binary-Concrete temperature \\
\bottomrule
\end{tabular}
\vspace{-2mm}
\end{table*}

\section{Related Work}
\label{sec:related}

\subsection{Stochastic Block Model}
The SBM~\cite{holland1983stochastic} is a probabilistic model for generating networks with community structures, where nodes are partitioned into blocks and edge probabilities depend on block memberships. Various methods have been developed for recovering community memberships from observed networks~\cite{abbe2018community}. Spectral clustering stands out for its computational tractability, and its statistical properties under SBM and its variant DCSBM have been extensively studied. For instance, weak consistency of clustering has been investigated by~\cite{rohe2011spectral, lei2015consistency}, among others, and strong consistency has been established by~\cite{su2019strong}. The minimax rate for estimating the edge probability matrix has also been characterized~\cite{gao2015rate}. The asymptotic Gaussian behavior of estimators for block probability matrices~\cite{tang2022asymptotically} and eigenvector matrices~\cite{tang2018limit, cape2019signal, xie2024entrywise} has been studied under SBMs, though the asymptotic behavior under more general DCSBMs remains less developed.
Recent work has extended SBMs to more complex settings. Mehta et al.~\cite{mehta2019stochastic} combine SBMs with GNN encoders for representation learning, though scalability remains limited. 
Li et al.~\cite{li2023ssbm} introduce a Signed Stochastic Block Model (SSBM) for multiple structure discovery in signed networks, highlighting its applicability to large-scale signed network analysis. Yang et al.~\cite{yang2025s3bm} propose a scalable Poisson-based S3BM for signed networks, focusing on degree correction and enhanced scalability.
Yang et al.~\cite{yang2025degree} propose degree-corrected SBMs for signed networks. These advances highlight the flexibility of SBM variants. 
In contrast to these works that focus primarily on community detection or representation learning, our work leverages OSBM within a neural variational framework specifically for link prediction, combining interpretable generative modeling with the expressiveness of neural encoders to scale to large-scale graphs.

\subsection{Link Prediction}
Link prediction aims to model the formation of links in graphs based on underlying factors~\cite{singh2024social}. We refer to these as ``LP factors'' and review two primary categories of methods: heuristics and MPNNs. We also discuss graph transformers, which have recently emerged as powerful alternatives.

\noindent\textbf{Heuristics for Link Prediction.}~~ Heuristic methods~\cite{newman2001clustering, zhou2009predicting} model LP factors via hand-crafted measures. Mao et al.~\cite{maorevisiting} identify three main factors: (1) local structural information, (2) global structural information, and (3) feature proximity. Local methods such as Common Neighbors (CN)~\cite{newman2001clustering}, Adamic Adar (AA)~\cite{adamic2003friends}, and Resource Allocation (RA)~\cite{zhou2009predicting} assume that nodes sharing more neighbors are likelier to connect. Global methods such as Katz~\cite{katz1953new} and Personalized PageRank (PPR)~\cite{brin1998anatomy} consider path-based similarities across the graph. Feature proximity methods~\cite{murase2019structural, nickel2014reducing, zhao2017leveraging} leverage node attributes. Beyond model design, Li et al.~\cite{li2023evaluating} identify pitfalls in existing LP benchmarks, particularly negative sampling bias, and propose the HeaRT protocol for more realistic evaluation, which we adopt in our experiments.

\noindent\textbf{MPNNs for Link Prediction.}~~Message Passing Neural Networks (MPNNs)~\cite{gilmer2017neural} learn node representations via message passing. Traditional MPNNs such as GCN~\cite{kipf2016semi}, SAGE~\cite{hamilton2017inductive}, and GAE~\cite{kipf2016variational} have been applied to link prediction but are suboptimal due to limited expressiveness for pairwise patterns~\cite{srinivasanequivalence, zhang2021labeling}. SEAL~\cite{zhang2018link} and NBFNet~\cite{zhu2021neural} address this by customizing message passing for each target link, enabling pairwise-specific learning. 
Zhao et al.~\cite{zhao2025grain} introduce GRAIN, a multi-granular and implicit information aggregation framework designed for heterophilous graphs. 
EGNN~\cite{zhao2025egnn} is a model that explores structure-level neighborhoods in graphs with varying homophily ratios, offering a promising direction for enhancing graph representation. However, these models incur quadratic complexity and do not scale well. 
Recent approaches~\cite{chamberlaingraph, wangneural, yun2021neo, zhao2025flexible} decouple message passing from pairwise computation to improve efficiency. For example, NCN/NCNC~\cite{wangneural} exploits common neighbor information, while BUDDY~\cite{chamberlaingraph} and Neo-GNN~\cite{yun2021neo} incorporate global structure.
Zhao et al.~\cite{zhao2025flexible} propose a flexible diffusion convolution method that improves the flexibility of graph neural networks.

\noindent\textbf{Graph Transformers.}~~Graph Transformers extend the Transformer architecture~\cite{vaswani2017attention} to graph-structured data. Graphormer~\cite{ying2021transformers} attends over all node pairs with structural, centrality, and edge encodings. SAN~\cite{kreuzer2021rethinking} enhances this with Laplacian positional encodings. TokenGT~\cite{kim2022pure} treats nodes and edges as tokens. 
To address this, sparse graph transformers~\cite{chennagphormer, wu2022nodeformer} have been proposed for node classification. For knowledge graph completion, some work~\cite{chen2021hitter, pahuja2023retrieve} has adapted transformers, but direct application to link prediction remains limited. LPFormer~\cite{shomer2024lpformer} introduces an adaptive graph transformer for LP but lacks the probabilistic interpretability of our TGSBM framework. Our work builds upon sparse graph transformers~\cite{shirzad2023exphormer} for efficiency and integrates them with OSBM's generative structure to provide both competitive predictive performance and interpretable community-based explanations, addressing a gap between the high performance of recent neural methods and the transparency required for practical deployment.

\section{Preliminaries}
\label{sec:preliminaries}

In this section, we introduce the fundamental concepts underlying TGSBM. We establish notation, formally define the link prediction task, and present the Overlapping Stochastic Block Model (OSBM)~\cite{latouche2011overlapping, miller2009nonparametric, zhu2016max} that provides the interpretable latent space for our approach. Table~\ref{tab:notation_compact} summarizes our notation.

\subsection{Problem Definition}

Let $G=(\mathcal{V},\mathcal{E})$ be a graph with $|\mathcal{V}|=N$ nodes and adjacency matrix $\mathbf{A}\in\{0,1\}^{N\times N}$. 
Given observed links and optional node features, link prediction aims to estimate the likelihood that a link exists between node pairs $(i,j)$ not in the observed training set. 
We denote a model's \emph{score} $s_\theta(i,j)\in\mathbb{R}$ and its corresponding link probability:
\begin{equation}
\hat{p}_{ij} = \sigma(s_\theta(i,j)), \qquad \sigma(x)=\frac{1}{1+e^{-x}}.
\label{eq:lp-core}
\end{equation}
This formulation encompasses similarity indices, embeddings, GNNs, and Transformer decoders, typically evaluated using binary classification metrics (AUC, Average Precision, MRR).

\begin{figure*}
  \centering
  \includegraphics[width=\linewidth]{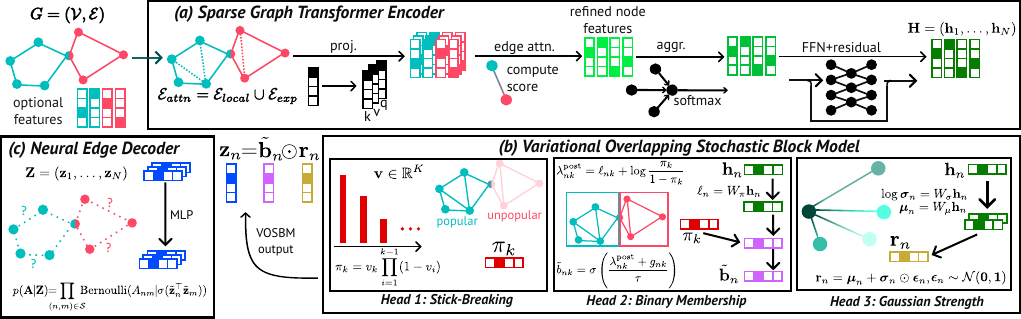}
  \caption{The overview of TGSBM architecture. TGSBM consists of three main components: (a) Sparse Graph Transformer Encoder processes node representations over local edges augmented with expander connections. (b) Variational Overlapping Stochastic Block Model uses three specialized heads to infer stick-breaking variables, binary community memberships, and Gaussian membership strengths. (c) Neural Edge Decoder reconstructs links via the OSBM generative process, maintaining interpretability through probabilistic community interactions.}
  \Description{Three-panel architecture diagram of TGSBM. Panel (a) shows a sparse graph transformer encoder with expander-augmented attention producing node embeddings. Panel (b) illustrates three variational heads inferring OSBM latent variables: stick-breaking for community prevalence (popular/unpopular communities shown as bars), binary membership via logit combination and relaxation, and Gaussian strength parameters. Panel (c) depicts a neural edge decoder with an MLP transformation and probabilistic link reconstruction.}
  \label{fig:overview}
\end{figure*}

\subsection{Overlapping Stochastic Block Model}

Real-world networks often exhibit \emph{overlapping} community structures, where nodes belong to multiple communities simultaneously. For instance, a researcher might be active in both machine learning and database communities. The Overlapping Stochastic Block Model (OSBM)~\cite{latouche2011overlapping, miller2009nonparametric, zhu2016max} provides a principled generative framework for such networks.

\noindent\textbf{Community memberships.}~~OSBM assumes each node may belong to multiple communities. Let $K$ denote the number (or truncation level) of latent communities, and $\mathbf{b}_i=(b_{i1},\ldots,b_{iK})\in\{0,1\}^K$ denote the \emph{binary} overlapping membership of node $i$. A common prior treats coordinates as independent Bernoulli variables:
\begin{equation}
p(\mathbf{b}_i\mid \boldsymbol{\alpha}) = \prod_{k=1}^{K} \alpha_k^{b_{ik}} (1-\alpha_k)^{1-b_{ik}},
\label{eq:osbm-membership}
\end{equation}
where $\boldsymbol{\alpha}=(\alpha_1,\ldots,\alpha_K)$ parameterizes community prevalence, and the all-zero vector is permitted to represent outliers.

\noindent\textbf{Edge generation.}~~Conditional on memberships, each edge $(i,j)$ follows a Bernoulli distribution with logistic natural parameter:
\begin{equation}
a_{ij} = \mathbf{b}_i^\top W\,\mathbf{b}_j + \mathbf{b}_i^\top U + V^\top \mathbf{b}_j + W_\ast,
\end{equation}
\begin{equation}
\Pr(A_{ij}=1\mid \mathbf{b}_i,\mathbf{b}_j)=\sigma(a_{ij}),
\label{eq:osbm-eta}
\end{equation}
where $W\in\mathbb{R}^{K\times K}$ captures inter-community and intra-community affinities, $U,V\in\mathbb{R}^{K}$ encode community-specific edge propensities, and $W_\ast\in\mathbb{R}$ provides global sparsity control. The block interaction term $\mathbf{b}_i^\top W\,\mathbf{b}_j$ promotes edges between nodes sharing communities, while the sender/receiver terms capture asymmetric community tendencies.

\noindent\textbf{Compact representation.}~~For notational convenience, we augment $\tilde{\mathbf{b}}_i=[\mathbf{b}_i;1]\in\{0,1\}^{K+1}$ and define $\tilde{W}=\begin{psmallmatrix} W & U \\ V^\top & W_\ast \end{psmallmatrix}$, yielding:
\begin{equation}
a_{ij} = \tilde{\mathbf{b}}_i^\top \tilde{W}\,\tilde{\mathbf{b}}_j.
\label{eq:osbm-aug}
\end{equation}

Under this formulation, edges are conditionally independent given latent memberships.

\noindent\textbf{Computational challenges.}~~While OSBM provides an elegant theoretical framework, inferring latent memberships $\mathbf{B}$ and community parameters from observed graphs poses significant computational challenges. Traditional approaches rely on variational EM or MCMC methods that scale poorly to large networks. TGSBM addresses these limitations by employing a sparse Graph Transformer encoder to efficiently infer variational posterior parameters while preserving OSBM's interpretable latent structure.

\section{TGSBM}
\label{sec:method}

TGSBM combines the interpretable latent structure of OSBM with the representational power of sparse Graph Transformers. The key insight is to preserve OSBM's principled generative model (Equations~\eqref{eq:osbm-membership}-\eqref{eq:osbm-aug}) while replacing traditional variational inference with a neural encoder that captures complex graph patterns through expander-augmented attention. Figure~\ref{fig:overview} provides an overview of the complete architecture. TGSBM consists of three main components: 
(1) A \emph{sparse Graph Transformer encoder} that processes node representations over local edges augmented with expander graph connections (Figure~\ref{fig:overview}(a));
(2) \emph{Variational heads} that output posterior parameters for OSBM latent variables: community memberships and strengths (Figure~\ref{fig:overview}(b));
(3) An \emph{edge decoder} that reconstructs links via the OSBM generative process (Figure~\ref{fig:overview}(c)).
The training objective combines edge reconstruction likelihood with KL regularization terms that encourage the learned representations to respect OSBM's hierarchical community structure. Algorithm~\ref{alg:tgsbm} summarizes the complete training procedure.

\subsection{Expander-Augmented Sparse Attention}
\label{subsec:sparse-attn}

Dense attention over all $O(N^2)$ node pairs is computationally prohibitive for large graphs. Following Exphormer~\cite{shirzad2023exphormer}, we construct a sparse attention topology that balances local structure preservation with efficient global information propagation through expander graphs, without requiring virtual nodes or global attention mechanisms (Figure~\ref{fig:overview}(a)).

\noindent\textbf{Attention edge construction.}~~Let $\mathcal{E}_{\text{local}}$ denote the original graph edges. We augment these with a $d_{\exp}$-regular expander graph $\mathcal{E}_{\exp}$ constructed via random permutations~\cite{friedman2003proof}: for each node $i$, we add edges $(i, \pi_j(i))$ where $\pi_1, \ldots, \pi_{d_{\exp}/2}$ are random permutations of $\{1,\ldots,N\}$, removing self-loops and duplicates. Following Theorem 4.3 in~\cite{shirzad2023exphormer}, this construction ensures $O(\log N)$ diameter with only $\Theta(d_{\exp}N)$ additional edges. The attention edge set becomes:
\begin{equation}
  \mathcal{E}_{\text{attn}} = \mathcal{E}_{\text{local}} \cup \mathcal{E}_{\exp}.
  \label{eq:attn-edges}
\end{equation}

\noindent\textbf{Theoretical justification.}~~The expander edges provide crucial theoretical guarantees for community detection:
\begin{itemize}[leftmargin=*]
\item \textbf{Spectral approximation}: The $d_{\exp}$-regular expander spectrally approximates the complete graph while maintaining a linear complexity~\cite{shirzad2023exphormer}.
\item \textbf{Rapid mixing}: Random walks mix in $O(\log N)$ steps, enabling efficient global information propagation without dense attention.
\item \textbf{Community preservation}: Unlike virtual nodes that create information bottlenecks, expander edges distribute connectivity uniformly, preserving local community structure while enabling long-range dependencies.
\end{itemize}

\noindent\textbf{Edge-type aware attention.}~~Given layer input $\mathbf{x}\in\mathbb{R}^{N\times D}$ with $H$ heads and per-head dimension $d_h$ ($D=H d_h$), we compute projections $Q(\mathbf{x}), K(\mathbf{x}), V(\mathbf{x})\in\mathbb{R}^{N\times H\times d_h}$. To distinguish between original and expander edges, we introduce learnable edge-type embeddings $E^{(h)}(\text{local}), E^{(h)}(\text{exp})\in\mathbb{R}^{d_h}$ for each head $h$, initialized randomly and learned during training.

For each edge $(u\to v)\in \mathcal{E}_{\text{attn}}$ and head $h$, attention scores combine content-based similarity with edge-type bias:
\begin{equation}
  s_{uv}^{(h)} = \frac{\langle Q^{(h)}(\mathbf{x}_v), K^{(h)}(\mathbf{x}_u)\rangle}{\sqrt{d_h}} + \langle Q^{(h)}(\mathbf{x}_v), E^{(h)}(\text{type}(u,v))\rangle,
  \label{eq:score}
\end{equation}
followed by softmax normalization and message aggregation:
\begin{equation}
  \alpha_{uv}^{(h)} = \frac{\exp(s_{uv}^{(h)})}{\sum_{u'\in\mathcal{N}(v)} \exp(s_{u'v}^{(h)})}, \quad
  \mathbf{m}_{v}^{(h)} = \sum_{u\in\mathcal{N}(v)} \alpha_{uv}^{(h)} V^{(h)}(\mathbf{x}_u),
  \label{eq:attn-agg}
\end{equation}
where $\mathcal{N}(v)=\{u:(u\to v)\in\mathcal{E}_{\text{attn}}\}$.

\noindent\textbf{Layer composition.}~~Head outputs are concatenated and processed through feed-forward networks with residual connections. We stack $L = O(\log N)$ layers to ensure full graph mixing~\cite{shirzad2023exphormer}, producing representations $\mathbf{h}_n\in\mathbb{R}^{D}$ that capture both local community structure and global patterns.

\noindent\textbf{Complexity analysis.}~~The computational cost per layer is $\mathcal{O}(ND^2 + |\mathcal{E}_{\text{attn}}|Hd_h) = \mathcal{O}((|\mathcal{E}|+d_{\exp}N)Hd_h)$, maintaining linear scaling in graph size compared to dense attention's $O(N^2)$ and standard GCN's $O(|\mathcal{E}|D^2)$.

\subsection{Variational Posterior and Reparameterization}
\label{subsec:variational}

Following variational approaches to OSBM inference~\cite{mehta2019stochastic}, we employ a structured variational family that respects the hierarchical nature of OSBM (Figure~\ref{fig:overview}(b)): global stick variables $\mathbf{v} = (v_1, \ldots, v_K) \in \mathbb{R}^K$ control community prevalence via stick-breaking construction~\cite{teh2007stick}, binary memberships $\mathbf{B} \in \{0,1\}^{N \times K}$ determine node-community assignments, and Gaussian strengths $\mathbf{R} \in \mathbb{R}^{N \times K}$ modulate membership intensities.

\noindent\textbf{Factorized variational family.}~~Our approximate posterior factorizes as:
\begin{align}
q(\mathbf{v}, \mathbf{B}, \mathbf{R} \mid \mathbf{X},\mathbf{A}) &= \prod_{k=1}^{K}\mathrm{Beta}(v_k \mid c_k,d_k) \nonumber \\ 
\quad \cdot \prod_{n=1}^{N}\prod_{k=1}^{K}\mathrm{Bern}(b_{nk} \mid \sigma(\ell_{nk})) &\quad \cdot \prod_{n=1}^{N}\mathcal{N}(\mathbf{r}_n \mid \boldsymbol{\mu}_n,\mathrm{diag}(\boldsymbol{\sigma}_n^2)),
\label{eq:variational-family}
\end{align}
where the encoder representations $\mathbf{H} = (\mathbf{h}_1, \ldots, \mathbf{h}_N) \in \mathbb{R}^{N \times D}$ parameterize distributions via specialized heads:
\begin{equation}
  \boldsymbol{\ell}_{n} = W_{\pi}\mathbf{h}_n, \quad 
  \boldsymbol{\mu}_n = W_{\mu}\mathbf{h}_n, \quad
  \log\boldsymbol{\sigma}_n = W_{\sigma}\mathbf{h}_n,
  \label{eq:heads}
\end{equation}
with projection matrices $W_\pi, W_\mu, W_\sigma \in \mathbb{R}^{K \times D}$.

\noindent\textbf{Stick-breaking integration.}~~Stick variables induce community prevalence priors via $v_k \sim \mathrm{Beta}(c_k, d_k)$ and cumulative probabilities $\pi_k = v_k \prod_{i=1}^{k-1} (1-v_i)$. The posterior membership logit combines encoder evidence with hierarchical priors:
\begin{equation}
  \lambda_{nk}^{\text{post}} = \ell_{nk} + \log\frac{\pi_k}{1-\pi_k}.
  \label{eq:post-membership-logit}
\end{equation}

\noindent\textbf{Reparameterized sampling.}~~For gradient-based optimization, we employ standard reparameterization techniques. Binary memberships use Binary-Concrete relaxation~\cite{jang2017categorical,maddison2017concrete}:
\begin{equation}
  \tilde{b}_{nk} = \sigma\left(\frac{\lambda_{nk}^{\text{post}} + g_{nk}}{\tau}\right), \quad g_{nk} \sim \mathrm{Gumbel}(0,1),
  \label{eq:binary-concrete}
\end{equation}
while continuous strengths follow Gaussian reparameterization:
\begin{equation}
  \mathbf{r}_n = \boldsymbol{\mu}_n + \boldsymbol{\sigma}_n \odot \boldsymbol{\epsilon}_n, \quad \boldsymbol{\epsilon}_n \sim \mathcal{N}(\mathbf{0},\mathbf{I}).
  \label{eq:gaussian-reparam}
\end{equation}
The final node embedding $\mathbf{z}_n=\tilde{\mathbf{b}}_n\odot\mathbf{r}_n$ ensures only active communities contribute to the representation, enabling low-variance path-wise gradients, where $\tilde{\mathbf{b}}_n$ denotes the relaxed binary membership vector from Equation~\eqref{eq:binary-concrete}.

\begin{table*}[t]
\centering
\caption{Dataset statistics. Split ratio denotes train/validation/test percentages.}
\label{tab:dataset_stats}
\vspace{-1mm}
\begin{tabular}{lccccc}
\toprule
 & \textbf{Cora} & \textbf{Citeseer} & \textbf{Pubmed} & \textbf{ogbl-collab} & \textbf{ogbl-ddi} \\
\midrule
\#Nodes & 2,708 & 3,312 & 19,717 & 235,868 & 4,267 \\
\#Edges & 5,278 & 4,552 & 44,324 & 1,285,465 & 1,334,889 \\
Mean Degree & 3.9 & 2.7 & 4.5 & 10.9 & 625.6 \\
Split Ratio & 85/5/10 & 85/5/10 & 85/5/10 & 92/4/4 & 80/10/10 \\
\midrule
Description & \multicolumn{3}{c}{Citation networks} & Collaboration network& Drug-drug interaction\\
\bottomrule
\end{tabular}
\vspace{-1mm}
\end{table*}

\subsection{Decoder and Likelihood}
\label{subsec:decoder}

The decoder translates OSBM's latent community embeddings $\mathbf{Z} = (\mathbf{z}_1, \ldots, \mathbf{z}_N) \in \mathbb{R}^{N \times K}$ into edge probabilities while preserving the interpretable generative structure (Figure~\ref{fig:overview}c). Following neural OSBM approaches~\cite{mehta2019stochastic}, we employ a learned nonlinear transformation to capture complex community interaction patterns beyond simple bilinear forms.

\noindent\textbf{Neural decoder architecture.}~~Edges are decoded via a two-layer MLP followed by an inner product and a logistic link:
\begin{equation}
f(\mathbf{z}) = \mathbf{W}_2\, \mathrm{ReLU}(\mathbf{W}_1 \mathbf{z} + \mathbf{b}_1) + \mathbf{b}_2, 
\qquad \tilde{\mathbf{z}}_n = f(\mathbf{z}_n),
\label{eq:mlp-decoder}
\end{equation}
where $\mathbf{W}_1 \in \mathbb{R}^{d_{\mathrm{hidden}} \times K}$, $\mathbf{W}_2 \in \mathbb{R}^{d_{\mathrm{mlp}} \times d_{\mathrm{hidden}}}$, e.g., $d_{\mathrm{hidden}} = 2K$ and $d_{\mathrm{mlp}} = K$.

The edge likelihood factorizes over evaluated pairs $\mathcal{S}$:
\begin{equation}
 p(\mathbf{A}\mid \mathbf{Z}) = \prod_{(n,m) \in \mathcal{S}} \mathrm{Bernoulli}(A_{nm} \mid \sigma(\tilde{\mathbf{z}}_n^\top \tilde{\mathbf{z}}_m)),
  \label{eq:edge-lhood-full}
\end{equation}
where $\mathcal{S}$ includes observed positive edges and uniformly sampled negatives for mini-batch training. This design maintains OSBM's symmetric structure while enabling flexible community interaction modeling.

\noindent\textbf{Optional feature reconstruction.}~~When node features $\mathbf{X}$ are available, we include reconstruction terms with feature-specific parameters, enabling joint edge-feature modeling for improved community discovery.

\subsection{Objective and Optimization}
\label{subsec:objective}

Following the variational autoencoder framework for OSBM~\cite{mehta2019stochastic}, we minimize the \emph{negative} of the evidence lower bound (ELBO) with OSBM priors $p(v_k)=\mathrm{Beta}(a,b)$, stick-breaking membership priors, and Gaussian strength priors $p(\mathbf{r})=\prod_n \mathcal{N}(\mathbf{0},\mathbf{I})$:
\begin{align}
 \mathcal{L} = \mathrm{KL}(q(\mathbf{v}) \| p(\mathbf{v})) + \mathrm{KL}(q(\mathbf{B}) \| p(\mathbf{B}\mid\mathbf{v})) \nonumber  \\ 
 \quad + \mathrm{KL}(q(\mathbf{R}) \| p(\mathbf{R})) - \mathbb{E}_{q}[\log p(\mathbf{A}\mid \mathbf{Z})] - \mathbb{E}_{q}[\log p(\mathbf{X}\mid \mathbf{Z})],
  \label{eq:elbo}
\end{align}
where $\mathrm{KL}(q(\cdot) \| p(\cdot))$ denotes the Kullback-Leibler divergence. The final term represents optional feature reconstruction when node features $\mathbf{X}$ are available; when features are absent, we omit this term. Links $\mathbf{A}$ are conditionally independent given the node embedding matrix $\mathbf{Z}$.

\begin{table*}[t]
\centering
\caption{Results on Cora, Citeseer, and Pubmed (\%) under standard evaluation. 
Highlighted are the results ranked \textcolor{myorange}{first}, \textcolor{mygreen}{second}, and \textcolor{myblue}{third}.}
\label{tab:results}
\vspace{-1mm}
\begin{tabular}{l l|c c|c c|c c}
\toprule
\multirow{2}{*}{} & \multirow{2}{*}{\textbf{Models}} 
 & \multicolumn{2}{c|}{\textbf{Cora}} 
 & \multicolumn{2}{c|}{\textbf{Citeseer}} 
 & \multicolumn{2}{c}{\textbf{Pubmed}} \\
 & & MRR & AUC & MRR & AUC & MRR & AUC \\
\midrule
\multirow{3}{*}{\textit{Heuristic}} 
  & CN & 20.99 & 70.85 & 28.34 & 67.49 & 14.02 & 63.9 \\
  & AA & 31.87 & 70.96 & 29.37 & 67.49 & 16.66 & 63.9 \\
  & RA & 30.79 & 70.96 & 27.61 & 67.48 & 15.63 & 63.9 \\
\midrule
\multirow{3}{*}{\textit{GNN}}
  & GCN & 32.50 $\pm$ 6.87 & 95.01 $\pm$ 0.32 & 50.01 $\pm$ 6.04 & 95.89 $\pm$ 0.26 & 19.94 $\pm$ 4.24 & 98.69 $\pm$ 0.06 \\
  & SAGE & \textcolor{myblue}{37.83 $\pm$ 7.75} & {95.63 $\pm$ 0.27} & 47.84 $\pm$ 6.39 & \textcolor{myblue}{97.39 $\pm$ 0.15} & 22.74 $\pm$ 5.47 & {98.87 $\pm$ 0.04} \\
  & GAE & 29.98 $\pm$ 3.21 & 95.08 $\pm$ 0.33 & \textcolor{myblue}{63.33 $\pm$ 3.14} & {97.06 $\pm$ 0.22} & 16.67 $\pm$ 0.19 & 97.47 $\pm$ 0.08 \\
\midrule
\multirow{6}{*}{\textit{GNN + Pairwise}}
  & SEAL & 26.69 $\pm$ 5.89 & 90.59 $\pm$ 0.75 & 39.36 $\pm$ 4.99 & 88.52 $\pm$ 1.40 & {38.06 $\pm$ 5.18} & 97.77 $\pm$ 0.40 \\
  & BUDDY & 26.40 $\pm$ 4.40 & 95.06 $\pm$ 0.36 & {59.48 $\pm$ 8.96} & 96.72 $\pm$ 0.26 & 23.98 $\pm$ 5.11 & 98.02 $\pm$ 0.05 \\
  & Neo-GNN & 22.65 $\pm$ 2.60 & 93.73 $\pm$ 0.36 & 53.97 $\pm$ 5.88 & 94.89 $\pm$ 0.60 & 32.33 $\pm$ 3.97 & 98.71 $\pm$ 0.05 \\
  & NCN & 32.93 $\pm$ 3.80 & \textcolor{myblue}{96.76 $\pm$ 0.18} & 54.97 $\pm$ 6.03 & 97.04 $\pm$ 0.26 & {35.65 $\pm$ 4.60} & \textcolor{myblue}{98.98 $\pm$ 0.04} \\
  & NCNC & 29.01 $\pm$ 3.83 & \textcolor{mygreen}{96.90 $\pm$ 0.28} & {64.03 $\pm$ 3.67} & \textcolor{myorange}{97.65 $\pm$ 0.30} & 25.70 $\pm$ 4.48 & \textcolor{mygreen}{99.14 $\pm$ 0.03} \\
  & NBFNet & {37.69 $\pm$ 3.97} & 92.85 $\pm$ 0.17 & 38.17 $\pm$ 3.06 & 91.06 $\pm$ 0.15 & \textcolor{myorange}{44.73 $\pm$ 2.12} & 98.34 $\pm$ 0.02 \\
\midrule
\multirow{2}{*}{\textit{Transformer}}
  & LPFormer & \textcolor{mygreen}{39.42 $\pm$ 5.78} & - & \textcolor{myorange}{65.42 $\pm$ 4.65} & - & \textcolor{myblue}{40.17 $\pm$ 1.92} & - \\
  & TGSBM (ours) & \textcolor{myorange}{39.69 $\pm$ 4.33} & \textcolor{myorange}{96.92 $\pm$ 0.19} & \textcolor{mygreen}{64.17 $\pm$ 4.06} & \textcolor{mygreen}{97.40 $\pm$ 0.19} & \textcolor{mygreen}{40.73 $\pm$ 1.12} & \textcolor{myorange}{99.34 $\pm$ 0.04} \\
\bottomrule
\end{tabular}
\vspace{-1mm}
\end{table*}

\noindent\textbf{Training procedure.}~~We employ stochastic gradient variational Bayes (SGVB)~\cite{kingma2013auto} with the reparameterization techniques from Equations~\eqref{eq:binary-concrete}-\eqref{eq:gaussian-reparam}. Beta stick variables are sampled via Kumaraswamy approximation~\cite{nalisnick2017stick}, and mini-batch edge sampling balances positive and negative pairs for reconstruction term estimation. The sparse attention design enables efficient scaling to large graphs while maintaining the interpretable OSBM latent structure. Algorithm~\ref{alg:tgsbm} provides the complete training procedure.

\section{Experiments}
\label{sec:experiments}

We conduct comprehensive experiments to evaluate TGSBM's performance, scalability, and interpretability. Our evaluation encompasses standard benchmarks and the challenging HeaRT protocol~\cite{li2023evaluating}, which employs realistic negative sampling to better reflect real-world deployment scenarios. We compare against a diverse set of baselines including heuristics, MPNNs, pairwise-enhanced models, and graph transformers, and provide ablation studies to validate our design choices.

\subsection{Experimental Settings}

\subsubsection{Datasets}
We evaluate TGSBM on five real-world datasets: three citation networks with node features and two large-scale collaboration graphs. For citation networks, all models receive partially-complete networks with an 85/5/10\% train/validation/test split. Table~\ref{tab:dataset_stats} provides detailed statistics for all datasets.

\textbf{Cora} is a citation network of scientific publications with sparse bag-of-words features. \textbf{Citeseer} contains publications from six research areas (agents, AI, databases, HCI, machine learning, information retrieval), with category labels one-hot encoded as node features. \textbf{Pubmed} consists of medical publications with bag-of-words features. \textbf{ogbl-collab} and \textbf{ogbl-ddi} are from the Open Graph Benchmark~\cite{hu2020open}, representing collaboration networks and drug-drug interactions, respectively. The latter has extremely high density (mean degree 625.6), posing significant scalability challenges.

\subsubsection{Baselines}
We compare against 13 representative methods spanning multiple paradigms: \textbf{Heuristics}--CN~\cite{newman2001clustering}, AA~\cite{adamic2003friends}, RA~\cite{zhou2009predicting}; \textbf{GNNs}--GCN~\cite{kipf2016semi}, SAGE~\cite{hamilton2017inductive}, GAE~\cite{kipf2016variational}; \textbf{Pairwise-enhanced GNNs}--SEAL~\cite{zhang2018link}, NBFNet~\cite{zhu2021neural}, Neo-GNN~\cite{yun2021neo}, BUDDY~\cite{chamberlaingraph}, NCN/NCNC~\cite{wangneural}; and \textbf{Graph Transformers}--LPFormer~\cite{shomer2024lpformer}. Results for citation networks follow Li et al.~\cite{li2023evaluating}, heuristics from Hu et al.~\cite{hu2020open}, and LPFormer from Shomer et al.~\cite{shomer2024lpformer}. Other results are from original publications or Chamberlain et al.~\cite{chamberlaingraph}.

\subsubsection{Hyperparameters} 
We tune learning rate in $\{10^{-3}, 5\times10^{-3}\}$, dropout in $[0, 0.3]$, and weight decay in $\{0, 10^{-5}, 10^{-4}\}$. Expander degree $d_{\exp}$ is searched over $\{6, 8, 10, 12\}$, number of layers in $\{3, 4, 5, 6\}$, with 4 attention heads. For the variational components, we use Beta$(10.0, 0.1)$ priors for stick-breaking, tune Concrete temperature in $\{0.8, 1.0, 1.2\}$ with prior temperature 0.5, and apply KL annealing from epoch 0 to $\{60, 80\}$ over 200 total epochs. Hidden dimension is 128 for Cora and Pubmed, 256 for Citeseer. We use GELU activation, BatchNorm, gradient clipping at 5.0, and report means over 3 random seeds.

\subsubsection{Evaluation Metrics} 
Each positive test edge is ranked against a set of negative candidates. We evaluate using three metrics: \textbf{AUC} (area under ROC curve), \textbf{MRR} (mean reciprocal rank), and \textbf{Hits@K} (proportion of correct links in top-K predictions). For Cora, Citeseer, and Pubmed, we follow Li et al.~\cite{li2023evaluating} and report AUC and MRR. For OGB datasets under HeaRT, we report MRR as the primary metric.

\subsection{Main Results}

Table~\ref{tab:results} presents performance on citation networks under standard evaluation. TGSBM achieves top-2 performance on 5 out of 6 (dataset, metric) pairs, demonstrating consistent strength across diverse graph structures.

\noindent\textbf{Per-dataset analysis.}~~On \textbf{Cora}, TGSBM achieves the best MRR (39.69) and AUC (96.92), outperforming LPFormer by 0.27 in MRR and all pairwise-enhanced methods in AUC. The improvements over pure GNN baselines are substantial: +7.2 MRR over SAGE and +1.9 AUC over NCNC. On \textbf{Citeseer}, TGSBM obtains second-best MRR (64.17), narrowly trailing LPFormer by 1.25, while achieving third-best AUC (97.20). Notably, NCNC leads in AUC (97.65), demonstrating the strength of common-neighbor features on this graph. On \textbf{Pubmed}, TGSBM secures second-best MRR (40.73), behind NBFNet's 44.73, but achieves the best AUC of 99.34 among all compared methods, surpassing NCNC's 99.14 by 0.20 points.

\noindent\textbf{Cross-method comparison.}~~Heuristics achieve moderate AUCs (63-71\%) but poor MRRs (14-32\%), confirming that topology-only scoring lacks discriminative power. Pure GNN encoders (GCN, SAGE, GAE) show high variance in MRR (standard deviations of 3-8), indicating instability in ranking quality. Pairwise-enhanced methods substantially improve both metrics, with NCNC and NBFNet reaching 44-64\% MRR, though at significant computational cost (separate inference per edge). Graph Transformers, particularly LPFormer and TGSBM, deliver the most consistent top-tier performance. Compared to LPFormer, TGSBM matches or exceeds MRR while additionally providing calibrated probability estimates (AUC), a critical property for production recommendation systems that require confidence scores.

\noindent\textbf{Key takeaways.}~~The results validate TGSBM's core design: expander-augmented sparse attention captures long-range dependencies efficiently, while the OSBM latent space models overlapping communities in a probabilistically principled manner. This combination yields robust performance across graphs of varying density (Cora: 0.0014, Citeseer: 0.0008, Pubmed: 0.0002 edge density) and scales (2.7K to 19.7K nodes).

\subsection{Performance on HeaRT Setting}
\label{subsec:heart}
\subsubsection{HeaRT evaluation protocol}
The HeaRT (Hard Negative Realistic Test) setting~\cite{li2023evaluating} addresses critical shortcomings in standard link prediction benchmarks. Traditional random negative sampling creates artificially easy evaluation scenarios where models can achieve high scores by exploiting spurious patterns. HeaRT instead constructs hard negatives via heuristic-based candidates (e.g., nodes with similar local structure but no actual connection), better reflecting real-world deployment where false positives are plausible non-edges. As Li et al.~\cite{li2023evaluating} demonstrate, this protocol causes substantial performance drops across all methods and can reverse relative model rankings, making it a more reliable indicator of production-ready performance.

\subsubsection{Results and analysis}
Table~\ref{tab:heart_mrr} reports MRR under HeaRT across five datasets. Absolute scores drop dramatically--e.g., from 65.42 to 9.90 on Pubmed for LPFormer--confirming the increased difficulty. TGSBM achieves a mean rank of 1.6 across all datasets, compared to LPFormer's 2.4, improving the average ranking position by 0.8.

\noindent\textbf{Method-wise trends.}~~Heuristics collapse to single-digit MRRs (mean rank 9.8-12.8), demonstrating complete failure against adversarial negatives. Pure GNN baselines show inconsistent performance: GAE achieves the best Cora score (18.32) but runs out of memory on ogbl-collab, while GCN obtains the second-best on ogbl-ddi (13.46) but lags on other datasets. Pairwise-enhanced methods are more robust, with NCN reaching the top Citeseer score (28.65), though NCNC's 24-hour timeout on ogbl-ddi highlights scalability limits. Graph Transformers dominate: LPFormer excels on Pubmed (9.90), while TGSBM achieves consistently top-2 performance on all five datasets, leading on Cora (16.98), Citeseer (26.92), ogbl-collab (7.77), and ogbl-ddi (13.73).

\noindent\textbf{TGSBM advantages.}~~TGSBM's consistent top-2 performance across all five datasets stems from two factors. First, expander-augmented attention provides efficient global mixing, crucial when hard negatives require distinguishing between nodes with similar local neighborhoods. Second, OSBM's overlapping community structure naturally captures the multiple affiliations present in collaboration networks (ogbl-collab) and complex interaction patterns in biomedical graphs (ogbl-ddi). The improved mean rank over LPFormer, combined with superior scalability (Section~\ref{subsec:runtime}), positions TGSBM as a practical solution for large-scale deployment.

\begin{table*}[t]
\centering
\caption{Results (MRR) under HeaRT evaluation. Highlighted are the results ranked \textcolor{myorange}{first}, \textcolor{mygreen}{second}, and \textcolor{myblue}{third}.}
\label{tab:heart_mrr}
\vspace{-1mm}
\begin{tabular}{ll|ccccc|c}
\toprule
 & \textbf{Models} &
\textbf{Cora} & \textbf{Citeseer} & \textbf{Pubmed} &
\textbf{ogbl-collab} & \textbf{ogbl-ddi} & \textbf{Mean Rank} \\
\midrule
\multirow{3}{*}{\textit{Heuristic}}&
CN  & 9.78  & 8.42  & 2.28  & 4.20 & 6.71  & 12.8 \\
& AA  & 11.91 & 10.82 & 2.63  & 5.07 & 6.97  & 11.0 \\
& RA  & 11.81 & 10.84 & 2.47  & 6.29 & 8.70  & 9.8 \\
\midrule
\multirow{3}{*}{\textit{GNN}}&
GCN  & 16.61 $\pm$ 0.30 & 21.09 $\pm$ 0.88 & 7.13 $\pm$ 0.27 & 6.09 $\pm$ 0.38 & \textcolor{mygreen}{13.46 $\pm$ 0.34}  & \textcolor{myblue}{5.0} \\
& SAGE & 14.74 $\pm$ 0.69 & 21.09 $\pm$ 1.15 & \textcolor{myblue}{9.40 $\pm$ 0.70} & 5.53 $\pm$ 0.50 & 12.60 $\pm$ 0.72 &  5.6 \\
& GAE  & \textcolor{myorange}{18.32 $\pm$ 0.41} & 25.25 $\pm$ 0.82 & 5.27 $\pm$ 0.25 & OOM & 3.49 $\pm$ 1.73 & NA \\
\midrule
\multirow{6}{*}{\textit{GNN + Pairwise}}
& SEAL    & 10.67 $\pm$ 3.46 & 13.16 $\pm$ 1.66 & 5.88 $\pm$ 0.53 & \textcolor{myblue}{6.43 $\pm$ 0.32} & 9.99 $\pm$ 0.90 & 8.6 \\
& BUDDY   & 13.71 $\pm$ 0.59 & 22.84 $\pm$ 0.36 & 7.56 $\pm$ 0.18 & 5.67 $\pm$ 0.36 &  12.43 $\pm$ 0.50  & 6.6 \\
& Neo-GNN & 13.95 $\pm$ 0.39 & 17.34 $\pm$ 0.84 & 7.74 $\pm$ 0.30 & 5.23 $\pm$ 0.90 &  10.86 $\pm$ 2.16  & 7.4 \\
& NCN     & 14.66 $\pm$ 0.95 & \textcolor{myorange}{28.65 $\pm$ 1.21} & 5.84 $\pm$ 0.22 & 5.09 $\pm$ 0.38 & 12.86 $\pm$ 0.78 & 6.0 \\
& NCNC    & 14.98 $\pm$ 1.00 & 24.10 $\pm$ 0.65 & 8.58 $\pm$ 0.59 &  4.73 $\pm$ 0.86 &  \,>24h\,  & 6.3 \\
& NBFNet  & 13.56 $\pm$ 0.58 & 14.29 $\pm$ 0.80 &  \,>24h\, & OOM &  \,>24h\, & NA \\
\midrule
\multirow{2}{*}{\textit{Transformer}}
 & LPFormer & \textcolor{myblue}{16.80 $\pm$ 0.52} & \textcolor{myblue}{26.34 $\pm$ 0.67} & \textcolor{myorange}{9.90 $\pm$ 0.52} & \textcolor{mygreen}{7.62 $\pm$ 0.26} & \textcolor{myblue}{13.20 $\pm$ 0.54} & \textcolor{mygreen}{2.4} \\
& TGSBM (ours) & \textcolor{mygreen}{16.98 $\pm$ 0.32} & \textcolor{mygreen}{26.92 $\pm$ 0.46} & \textcolor{mygreen}{9.61 $\pm$ 0.53} & \textcolor{myorange}{7.77 $\pm$ 0.29} & \textcolor{myorange}{13.73 $\pm$ 0.42} & \textcolor{myorange}{1.6} \\
\bottomrule
\end{tabular}
\vspace{-1mm}
\end{table*}

\subsection{Runtime Analysis}
\label{subsec:runtime}

\begin{figure}[t]
  \centering
  \includegraphics[width=\linewidth]{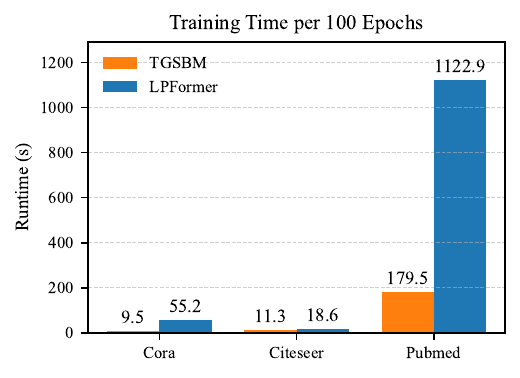}
  \caption{Training time per 100 epochs on Cora, Citeseer, and Pubmed. TGSBM (orange) vs. LPFormer (blue). Bars represent median wall-clock time across 3 runs.}
  \Description{Bar chart comparing training time: TGSBM takes 9.5s, 11.3s, 179.5s on Cora, Citeseer, Pubmed; LPFormer takes 55.2s, 18.6s, 1122.9s respectively, showing TGSBM's 6× speedup on the largest graph.}
  \label{fig:train_time_epoch}
  \vspace{-3mm}
\end{figure}

We compare TGSBM's training efficiency against LPFormer, the strongest-performing baseline. Figure~\ref{fig:train_time_epoch} reports wall-clock time per 100 epochs on Cora, Citeseer, and Pubmed, measured on a single NVIDIA A100 GPU with identical batch sizes.

On small graphs (Cora and Citeseer), TGSBM requires 9.5s and 11.3s compared to LPFormer's 55.2s and 18.6s, yielding 5.8$\times$ and 1.6$\times$ speedups respectively. The efficiency gap widens dramatically on Pubmed: TGSBM finishes in 179.5s while LPFormer takes 1122.9s, a 6.3$\times$ speedup. This scaling behavior directly reflects architectural differences. TGSBM's complexity is $O((|E|+d_{\exp}N)Hd_h)$, linear in graph size. In contrast, LPFormer employs PPR-based $k$-hop neighborhood expansion followed by pairwise attention, incurring $O(k|E| + N_{\text{PPR}}^2)$ cost where $N_{\text{PPR}}$ can exceed $O(N)$ on dense graphs.

Projecting to larger OGB benchmarks: ogbl-collab (1.28M edges) and ogbl-citation2 (30M edges) would incur even more severe penalties for quadratic attention mechanisms. The extremely dense ogbl-ddi (mean degree 625) and ogbl-ppa (mean degree 100+) further exacerbate this gap. Our results demonstrate that TGSBM achieves competitive or superior accuracy while maintaining practical scalability for large-scale graphs--a critical requirement for production deployment where training time directly impacts iteration cycles and cost.

\section{Conclusion}
\label{sec:conclusion}

We introduced TGSBM, a framework that integrates the interpretable latent structure of Overlapping Stochastic Block Models with the scalability of sparse Graph Transformers for large-scale link prediction. Through expander-augmented sparse attention, neural variational inference, and a generative edge decoder, TGSBM addresses important challenges in modeling large-scale networks: efficient global reasoning, probabilistic capture of overlapping communities, and interpretable prediction explanations.
Our evaluation demonstrates that TGSBM achieves competitive performance across diverse benchmarks, obtaining top-2 rankings on 8 out of 10 dataset-metric pairs and attaining a mean rank of 1.6 under the challenging HeaRT protocol, improving upon the strongest Graph Transformer baseline. Beyond predictive accuracy, TGSBM delivers up to 6$\times$ faster training while maintaining interpretable community structures. Ablation studies confirm that each component contributes meaningfully to this robust performance.
TGSBM's integration of probabilistic modeling and neural architectures opens avenues for extensions to dynamic graphs and multi-modal networks. More broadly, our work demonstrates that interpretability and scalability can be effectively combined: by grounding neural graph learning in well-established generative models, we can build systems that balance accuracy, efficiency, and transparency--properties that are valuable for large-scale deployment in real-world applications.

\begin{acks}
This work is supported in part by the Key R\&D Project of Jilin Province (No. 20240304200SF).
\end{acks}

\newpage

\appendix

\begin{figure*}[t]
\centering
\includegraphics[width=\linewidth]{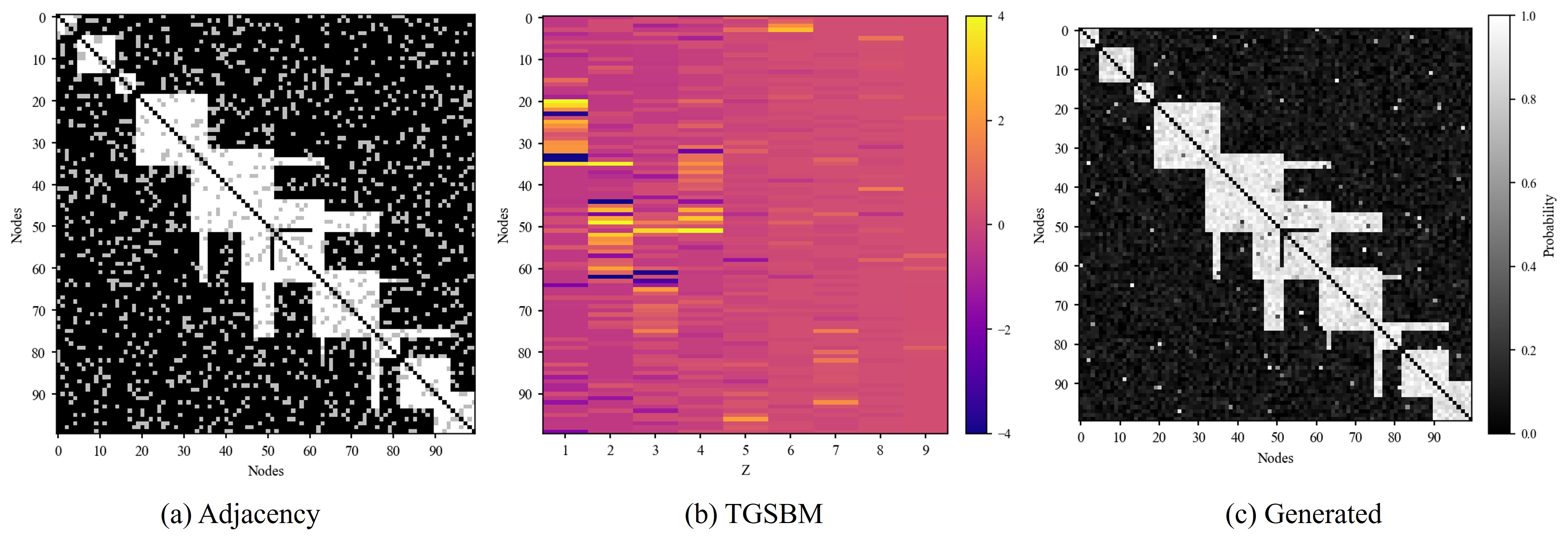}
\caption{Qualitative analysis on synthetic data with 100 nodes and 10 communities. (a) Ground-truth adjacency matrix sorted by community memberships, with white/black/gray denoting link/no-link/hidden edges. (b) Learned latent structure $\mathbf{Z} \in \mathbb{R}^{100 \times 10}$ from TGSBM, where color intensity indicates community membership strength. (c) Reconstructed adjacency probabilities using only the first 2 dimensions of $\mathbf{Z}$ (columns 1-2), demonstrating that dominant communities capture essential network structure.}
\label{fig:synthetic}
\end{figure*}




\section{Pseduocode}

\begin{algorithm}[h]
\caption{TGSBM Training Algorithm}
\label{alg:tgsbm}
\begin{algorithmic}[1]
\REQUIRE Graph $G=(\mathcal{V}, \mathcal{E})$, node features $\mathbf{X}$ (optional), hyperparameters $\{K, H, L, d_{\exp}, \tau\}$
\ENSURE Trained TGSBM parameters $\{\theta, \phi\}$
\STATE \textbf{Initialize:} Transformer parameters $\theta$, variational heads $\phi$, edge-type embeddings
\STATE \textbf{Construct sparse attention topology:}
\STATE \quad $\mathcal{E}_{\text{local}} \leftarrow \mathcal{E}$
\STATE \quad $\mathcal{E}_{\exp} \leftarrow$ Generate $d_{\exp}$-regular expander via random permutations
\STATE \quad $\mathcal{E}_{\text{attn}} \leftarrow \mathcal{E}_{\text{local}} \cup \mathcal{E}_{\exp}$
\WHILE{not converged}
    \STATE \textbf{// Sparse Graph Transformer Encoder}
    \STATE $\mathbf{h}^{(0)} \leftarrow \mathbf{X}$ (or random initialization if no features)
    \FOR{$\ell = 1$ to $L$}
        \STATE Compute edge-type aware attention scores via Eq.~\eqref{eq:score}
        \STATE $\mathbf{h}^{(\ell)} \leftarrow \text{TransformerLayer}(\mathbf{h}^{(\ell-1)}, \mathcal{E}_{\text{attn}})$
    \ENDFOR
    
    \STATE \textbf{// Variational Posterior Parameters}
    \STATE $\boldsymbol{\ell}_n \leftarrow W_\pi \mathbf{h}_n^{(L)}$, $\boldsymbol{\mu}_n \leftarrow W_\mu \mathbf{h}_n^{(L)}$, $\log\boldsymbol{\sigma}_n \leftarrow W_\sigma \mathbf{h}_n^{(L)}$
    
    \STATE \textbf{// Reparameterized Sampling}
    \STATE Sample stick variables $v_k \sim \text{Beta}(c_k, d_k)$ for $k = 1, \ldots, K$
    \STATE Compute stick-breaking priors $\lambda_k^{\text{prior}} \leftarrow \log\frac{\pi_k}{1-\pi_k}$
    \STATE $\lambda_{nk}^{\text{post}} \leftarrow \ell_{nk} + \lambda_k^{\text{prior}}$
    \STATE $\tilde{b}_{nk} \leftarrow \sigma\left(\frac{\lambda_{nk}^{\text{post}} + g_{nk}}{\tau}\right)$ where $g_{nk} \sim \text{Gumbel}(0,1)$
    \STATE $r_{nk} \leftarrow \mu_{nk} + \sigma_{nk} \epsilon_{nk}$ where $\epsilon_{nk} \sim \mathcal{N}(0,1)$
    \STATE $\mathbf{z}_n \leftarrow \tilde{\mathbf{b}}_n \odot \mathbf{r}_n \in \mathbb{R}^K$
    
    \STATE \textbf{// Neural Edge Decoder}
    \STATE $\tilde{\mathbf{z}}_n \leftarrow f(\mathbf{z}_n)$ via two-layer MLP (Eq.~\eqref{eq:mlp-decoder})
    \STATE Sample edge pairs $\mathcal{S}$ (positive + negative sampling)
    
    \STATE \textbf{// Negative ELBO Computation}
    \STATE $\mathcal{L}_{\text{recon}} \leftarrow \sum_{(n,m) \in \mathcal{S}} A_{nm}\log\sigma(\tilde{\mathbf{z}}_n^\top\tilde{\mathbf{z}}_m) + (1-A_{nm})\log(1-\sigma(\tilde{\mathbf{z}}_n^\top\tilde{\mathbf{z}}_m))$
    \STATE $\mathcal{L}_{\mathrm{KL}} \leftarrow \mathrm{KL}(q(\mathbf{v}) \| p(\mathbf{v})) + \mathrm{KL}(q(\mathbf{B}) \| p(\mathbf{B}\mid\mathbf{v})) + \mathrm{KL}(q(\mathbf{R}) \| p(\mathbf{R}))$
    \STATE $\mathcal{L} \leftarrow  \mathcal{L}_{\mathrm{KL}} - \mathcal{L}_{\text{recon}}$
    \STATE Update parameters $\theta, \phi$ via $\nabla_{\theta,\phi} \mathcal{L}$
\ENDWHILE
\end{algorithmic}
\end{algorithm}

\section{Time Complexity Analysis}

We provide a detailed derivation of TGSBM's time complexity, focusing on the sparse Graph Transformer encoder that dominates the computational cost. We show that TGSBM achieves near-linear complexity $O((|\mathcal{E}| + d_{\exp}N)Hd_h)$, making it scalable to Web-sized graphs.

\subsection{Sparse Graph Transformer Complexity}

The encoder consists of $L$ transformer layers operating on the sparse attention topology $\mathcal{E}_{\text{attn}} = \mathcal{E}_{\text{local}} \cup \mathcal{E}_{\text{exp}}$, where $|\mathcal{E}_{\text{local}}| = |\mathcal{E}|$ and $|\mathcal{E}_{\text{exp}}| = d_{\exp}N$. Thus:
\begin{equation}
|\mathcal{E}_{\text{attn}}| = |\mathcal{E}| + d_{\exp}N
\end{equation}

\subsubsection{Per-Layer Operations}

For each layer $\ell \in [L]$, the computation proceeds as follows:

\paragraph{Step 1: Linear projections.} 
Given input $\mathbf{h}^{(\ell-1)} \in \mathbb{R}^{N \times D}$, compute Query, Key, Value:
\begin{equation}
Q = W_Q \mathbf{h}^{(\ell-1)}, \quad K = W_K \mathbf{h}^{(\ell-1)}, \quad V = W_V \mathbf{h}^{(\ell-1)}
\end{equation}
where $W_Q, W_K, W_V \in \mathbb{R}^{D \times D}$.

\noindent\textbf{Cost:}~~$O(ND^2)$ for three matrix multiplications.

\paragraph{Step 2: Sparse attention computation.}
For each edge $(u \to v) \in \mathcal{E}_{\text{attn}}$ and head $h \in [H]$, compute attention scores (Equation 7 in main text):
\begin{equation}
s_{uv}^{(h)} = \frac{\langle Q_v^{(h)}, K_u^{(h)} \rangle}{\sqrt{d_h}} + \langle Q_v^{(h)}, E^{(h)}_{\text{type}(u,v)} \rangle
\end{equation}
where $d_h = D/H$ is the per-head dimension.

\noindent\textbf{Cost per edge:}~~$O(Hd_h)$ for computing $H$ inner products of dimension $d_h$.

\noindent\textbf{Total over all edges:}~~$O(|\mathcal{E}_{\text{attn}}| \cdot Hd_h)$.

\paragraph{Step 3: Softmax and aggregation.}
Normalize attention weights and aggregate messages:
\begin{align}
\alpha_{uv}^{(h)} &= \frac{\exp(s_{uv}^{(h)})}{\sum_{u' \in \mathcal{N}(v)} \exp(s_{u'v}^{(h)})}, \\
\mathbf{m}_v^{(h)} &= \sum_{u \in \mathcal{N}(v)} \alpha_{uv}^{(h)} V_u^{(h)}
\end{align}

\noindent\textbf{Cost:}~~$O(|\mathcal{E}_{\text{attn}}| \cdot Hd_h)$ for weighted sum over edges.

\paragraph{Step 4: Feed-forward network.}
Concatenate head outputs and apply FFN:
\begin{equation}
\mathbf{h}^{(\ell)} = \text{FFN}(\text{Concat}(\mathbf{m}^{(1)}, \ldots, \mathbf{m}^{(H)}))
\end{equation}

\noindent\textbf{Cost:}~~$O(ND^2)$ for two-layer MLP with hidden dimension $4D$ (standard Transformer configuration).

\paragraph{Per-layer total.}
Combining Steps 1-4:
\begin{align}
T_{\text{layer}} &= O(ND^2) + O(|\mathcal{E}_{\text{attn}}| \cdot Hd_h) + O(|\mathcal{E}_{\text{attn}}| \cdot Hd_h) + O(ND^2) \\
&= O(ND^2 + |\mathcal{E}_{\text{attn}}| \cdot Hd_h)
\end{align}

For large sparse graphs, the attention term dominates: $|\mathcal{E}_{\text{attn}}| \cdot Hd_h \gg ND^2$, since $|\mathcal{E}_{\text{attn}}| = \Omega(N)$ and $D = Hd_h$. Therefore:
\begin{equation}
T_{\text{layer}} = O(|\mathcal{E}_{\text{attn}}| \cdot Hd_h) = O((|\mathcal{E}| + d_{\exp}N) \cdot Hd_h)
\end{equation}

\subsubsection{Full Encoder}

With $L$ layers, the total encoder cost is:
\begin{equation}
T_{\text{encoder}} = L \cdot T_{\text{layer}} = O(L \cdot (|\mathcal{E}| + d_{\exp}N) \cdot Hd_h)
\end{equation}

Following Exphormer \cite{shirzad2023exphormer}, $L = O(\log N)$ layers suffice for global information propagation due to the expander's $O(\log N)$ diameter. In practice, $L \in [3, 6]$ for graphs with $N \in [10^3, 10^6]$, so we treat $L$ as a small constant. Thus:
\begin{equation}
\boxed{T_{\text{encoder}} = O((|\mathcal{E}| + d_{\exp}N) \cdot Hd_h)}
\end{equation}

\subsection{Other Components}

The variational heads, sampling, and decoder contribute lower-order terms:
\begin{itemize}
\item \textbf{Variational heads:} $O(NKD)$ for linear projections, where $K \ll D$.
\item \textbf{Reparameterized sampling:} $O(NK)$ for Binary-Concrete and Gaussian sampling.
\item \textbf{Decoder:} $O(NK^2)$ for MLP transformation and $O(BK)$ for edge probability computation over mini-batch $\mathcal{S}$ of size $B$.
\item \textbf{ELBO:} $O(NK + B)$ for KL terms and reconstruction likelihood.
\end{itemize}

Since $K \ll D$ and the encoder dominates, the overall TGSBM complexity is:
\begin{equation}
\boxed{T_{\text{TGSBM}} = O((|\mathcal{E}| + d_{\exp}N) \cdot Hd_h)}
\end{equation}

For sparse graphs where $|\mathcal{E}| = \Theta(N)$, and with constant $d_{\exp} \in [6, 12]$:
\begin{equation}
T_{\text{TGSBM}} = O(N \cdot Hd_h) = O(ND)
\end{equation}
which is linear in graph size.

\subsection{Comparison with Dense Transformers}

Dense Graph Transformers (e.g., Graphormer \cite{ying2021transformers}) compute attention over all $N^2$ node pairs:
\begin{equation}
T_{\text{dense}} = O(LN^2D)
\end{equation}

\noindent\textbf{Speedup factor:}~~
\begin{equation}
\frac{T_{\text{dense}}}{T_{\text{TGSBM}}} = \frac{N^2D}{(|\mathcal{E}| + d_{\exp}N) \cdot Hd_h} \approx \frac{N^2}{|\mathcal{E}|} = \frac{1}{\text{graph density}}
\end{equation}

For citation networks with typical density $\rho \in [0.0002, 0.002]$, the speedup is proportional to $\frac{1}{\rho}$, providing orders of magnitude reduction in attention operations.

\subsection{Summary}

TGSBM achieves near-linear time complexity $O((|\mathcal{E}| + d_{\exp}N) \cdot Hd_h)$ through:
\begin{enumerate}[leftmargin=*]
\item \textbf{Sparse attention:} Computing attention only over $|\mathcal{E}_{\text{attn}}| = O(|\mathcal{E}| + N)$ edges instead of all $N^2$ pairs.
\item \textbf{Expander augmentation:} Adding $d_{\exp}N$ edges with $d_{\exp} \ll N$ to ensure $O(\log N)$ diameter without quadratic cost.
\item \textbf{Efficient attention:} Per-head dimension $d_h = D/H$ reduces computation compared to full-dimensional inner products.
\end{enumerate}

This complexity profile makes TGSBM scalable to Web-sized graphs with millions of nodes, as confirmed empirically in Section~\ref{subsec:runtime} where TGSBM trains up to $6\times$ faster than LPFormer on medium-scale benchmarks.

\section{Qualitative Analysis of Learned Embeddings}

To demonstrate the interpretable nature of embeddings learned by TGSBM, we analyze a synthetic dataset with known community structure and visualize the learned latent representations.

\subsection{Synthetic Data}

To demonstrate the interpretable nature of the embeddings learned by our model, we generate a synthetic dataset with 100 nodes and 10 communities. The dataset is generated by fixing the ground-truth communities (by creating a binary vector for each node) such that some of the nodes belong to the same communities. The adjacency matrix is then generated using a simple inner product, followed by the sigmoid operation (Figure~\ref{fig:synthetic}(a)). We train using 85\% of the synthetic adjacency matrix for link-prediction and for visualizing the latent structure that our model learns.

\subsection{Learned Community Structure}

Figure~\ref{fig:synthetic}(b) shows the learned latent structure $\mathbf{Z} \in \mathbb{R}^{N \times K}$. The visualization reveals several key properties:

\noindent\textbf{Overlapping memberships.}~~Nodes exhibit interpretable overlapping patterns. For example, nodes 20-30 show high membership strengths in both Communities 1 and 2 (columns 0-1), reflecting their multi-community roles. This demonstrates TGSBM's ability to capture nodes belonging to multiple groups simultaneously.

\noindent\textbf{Stick-breaking effect.}~~The stick-breaking prior encourages commonly-selected communities to concentrate in earlier columns (left side of the heat map), while less-prevalent communities appear later with sparser assignments. This ordering enables automatic determination of the effective number of communities.

\noindent\textbf{Sparse interpretability.}~~Despite continuous relaxation during training, the posterior concentrates on sparse patterns where most nodes have strong memberships in only 1-3 communities, facilitating clear community assignments.

\subsection{Low-Dimensional Reconstruction}

Figure~\ref{fig:synthetic}(c) demonstrates that TGSBM's community structure provides a compact representation. Using only the \textit{first two dimensions} of $\mathbf{Z}$ (columns 0-1 in Figure~\ref{fig:synthetic}(b)), with all other dimensions set to zero, we reconstruct the adjacency matrix via the decoder. The reconstructed graph closely matches the original, showing that:

\begin{itemize}
\item The dominant communities capture essential network structure
\item The stick-breaking prior successfully identifies that only a subset of $K=10$ communities are necessary
\item Link formation can be explained via shared community memberships--e.g., strong connections between nodes 1-20 arise from their shared membership in Community 1
\end{itemize}

This property is crucial for interpretability: practitioners can focus on the top few communities to understand and explain link predictions, rather than examining all $K$ dimensions.

\subsection{Summary}

The qualitative analysis confirms that TGSBM learns interpretable overlapping community structures that:
(i) recover ground-truth overlapping memberships,
(ii) produce sparse representations amenable to human interpretation,
(iii) automatically determine effective dimensionality, and
(iv) provide transparent explanations for link predictions via community affiliations.
These properties validate TGSBM's combination of OSBM interpretability with neural network expressiveness, which is essential for Web applications requiring both accuracy and explainability.

\subsection{Ablation Study}

To isolate the contribution of each architectural component, we conduct ablation studies on Citeseer and Pubmed by progressively removing: (i) expander edges (\textbf{w/o Expander}), (ii) local edges (\textbf{w/o Local}), (iii) stick-breaking priors (\textbf{w/o Stick-breaking prior}), (iv) KL regularization on binary memberships (\textbf{w/o KL on memberships}), and (v) KL regularization on Gaussian strengths (\textbf{w/o KL on strengths}).

Table~\ref{tab:ablation_lpformer} reports AUC performance. On Citeseer, removing expander edges drops AUC from 97.20 to 95.42 ($-1.78$), demonstrating that global information propagation is critical for this medium-density graph. Removing local edges yields 94.86 ($-2.34$), confirming that expander shortcuts alone cannot recover fine-grained community structure. Eliminating the stick-breaking prior reduces AUC to 95.77 ($-1.43$), indicating that hierarchical community modeling provides stable inductive bias. Removing KL terms on memberships or strengths also degrades performance (95.11 and 96.02 respectively), as these regularizers prevent overfitting to the observed graph structure.

On Pubmed, trends are consistent but attenuated: w/o Expander achieves 98.62 ($-0.72$), w/o Local 98.11 ($-1.23$), w/o Stick-breaking prior 98.49 ($-0.85$), w/o KL on memberships 98.23 ($-1.11$), and w/o KL on strengths 98.72 ($-0.62$). The smaller magnitude reflects Pubmed's higher density (44K edges vs. 4-5K), where local structure provides stronger signals. Nevertheless, the consistent degradation across all ablations confirms that each component contributes meaningfully to TGSBM's performance.

\begin{table}[t]
\centering
\caption{Ablation study on TGSBM components (AUC).}
\label{tab:ablation_lpformer}
\vspace{-1mm}
\begin{tabular}{l|c|c}
\toprule
\textbf{Method} & \textbf{Citeseer} & \textbf{Pubmed} \\
\midrule
w/o Expander            & 95.42 $\pm$ 0.27 & 98.62 $\pm$ 0.05 \\
w/o Local               & 94.86 $\pm$ 0.31 & 98.11 $\pm$ 0.07 \\
w/o Stick-breaking prior& 95.77 $\pm$ 0.29 & 98.49 $\pm$ 0.06 \\
w/o KL on memberships   & 95.11 $\pm$ 0.33 & 98.23 $\pm$ 0.08 \\
w/o KL on strengths     & 96.02 $\pm$ 0.25 & 98.72 $\pm$ 0.04 \\
\midrule
TGSBM (full)            & $\mathbf{97.20{\pm}0.19}$ & $\mathbf{99.34{\pm}0.04}$ \\
\bottomrule
\end{tabular}
\end{table}

\end{document}